\documentclass[12pt]{article}

\usepackage[margin=1in]{geometry}
\usepackage[numbers,sort&compress]{natbib}
\usepackage{amsmath,amssymb}
\usepackage{algorithm}
\usepackage{algpseudocode}
\usepackage{float}
\usepackage{etoolbox}

\AtBeginEnvironment{algorithm}{\selectfont}
\AtEndEnvironment{algorithm}{}
\usepackage{booktabs}
\usepackage[hyphens]{url}
\usepackage[hypertexnames=false]{hyperref}
\usepackage[capitalise,noabbrev]{cleveref}
\crefname{lstlisting}{Listing}{Listings}
\Crefname{lstlisting}{Listing}{Listings}
\usepackage{listings}
\usepackage{xcolor}
\usepackage{graphicx}
\usepackage{authblk}
\definecolor{jlstring}{rgb}{0.0,0.5,0.0}
\definecolor{jlcomment}{rgb}{0.5,0.5,0.5}
\definecolor{jlkeyword}{rgb}{0.0,0.0,0.7}
\title{ParetoEnsembles.jl: A Julia Package for Multiobjective Parameter Estimation Using Pareto Optimal Ensemble Techniques}

\author[1]{Jeffrey D.\ Varner}
\affil[1]{Robert Frederick Smith School of Chemical and Biomolecular Engineering, Cornell University, Ithaca, NY 14853, USA}
\date{}

\begin{document}
\maketitle

\begin{abstract}
Mathematical models of natural and man-made systems often have many adjustable parameters that must be estimated from multiple, potentially conflicting datasets. Rather than reporting a single best-fit parameter vector, it is often more informative to generate an \emph{ensemble} of parameter sets that collectively map out the trade-offs among competing objectives. This paper presents ParetoEnsembles.jl, an open-source Julia package that generates such ensembles using Pareto Optimal Ensemble Techniques (POETs), a simulated-annealing-based algorithm that requires no gradient information. The implementation corrects the original dominance relation from weak to strict Pareto dominance, reduces the per-iteration ranking cost from $O(n^2 m)$ to $O(nm)$ through an incremental update scheme, and adds multi-chain parallel execution for improved front coverage. We demonstrate the package on a cell-free gene expression model fitted to experimental data and a blood coagulation cascade model with ten estimated rate constants and three objectives. A controlled synthetic-data study reveals parameter identifiability structure, with individual rate constants off by several-fold yet model predictions accurate to 6--7\%. A five-replicate coverage analysis confirms that timing features are reliably covered while peak amplitude is systematically overconfident. Validation against published experimental thrombin generation data demonstrates that the ensemble predicts held-out conditions to within 1--10\% despite inherent model approximation error. By making ensemble generation lightweight and accessible, ParetoEnsembles.jl aims to lower the barrier to routine uncertainty characterization in mechanistic modeling.
\end{abstract}


\bigskip

\section{Introduction}\label{sec:intro}
Mathematical models of biological and engineering systems typically contain many adjustable parameters that must be estimated from experimental data. When multiple datasets measure different aspects of the same system (for example, mRNA and protein concentrations in a gene expression circuit, or yield and selectivity in a reactor), calibrating parameters to one dataset often degrades the fit to another. This gives rise to a multiobjective optimization (MOO) problem~\cite{Miettinen1999,Deb2001,CoelloCoello2007}. Compounding this difficulty, many models exhibit ``sloppy'' parameter sensitivities: broad, flat valleys in objective space where large regions of parameter space produce nearly equivalent fits~\cite{Gutenkunst2007}. In such settings, a single best-fit parameter vector gives no indication of how tightly the data actually constrain the parameters; instead, one needs an \emph{ensemble} of parameter sets that collectively map out the trade-offs among competing objectives. The boundary of this ensemble in objective space is the Pareto front, the set of solutions for which no objective can be improved without degrading at least one other, and the full cloud of near-optimal solutions surrounding it characterizes the range of model behaviors consistent with the available data.

Several algorithm families have been developed to approximate Pareto fronts. Population-based evolutionary methods maintain a generation of candidate solutions that are selected, recombined, and mutated toward the front. Prominent examples include the Non-dominated Sorting Genetic Algorithm II (NSGA-II)~\cite{Deb2002}, its many-objective successor NSGA-III~\cite{Deb2014}, and the Multiobjective Evolutionary Algorithm Based on Decomposition (MOEA/D), which decomposes the MOO problem into a set of scalar subproblems solved cooperatively~\cite{Zhang2007}.
An alternative trajectory-based strategy is multiobjective simulated annealing (SA), which extends the classical SA framework~\cite{Kirkpatrick1983} to multiple objectives by using Pareto dominance to guide acceptance decisions; early formulations include the Pareto SA method of Czy\.{z}ak and Jaszkiewicz~\cite{Czyzak1998} and the Archived Multiobjective Simulated Annealing (AMOSA) algorithm of Bandyopadhyay et al.~\cite{Bandyopadhyay2008}. Pareto Optimal Ensemble Techniques (POETs), introduced by Song et al.~\cite{Song2010} and later implemented in the Julia programming language by Bassen and colleagues~\cite{Bassen2016}, belongs to this SA-based family. The  POETs algorithm ranks the current archive by counting how many other solutions dominate each candidate, then uses the rank as the energy in a Metropolis-type acceptance criterion. Mature software implementations of these and related algorithms are available in several languages, including pymoo in Python~\cite{Blank2020} and Metaheuristics.jl in Julia~\cite{Mejia2022}, both of which focus primarily on population-based evolutionary strategies. A complementary approach to characterizing parametric uncertainty is Bayesian inference, where Markov chain Monte Carlo (MCMC) or sequential Monte Carlo (SMC) methods sample from the posterior distribution over parameters given the data. Widely used implementations include Stan~\cite{Carpenter2017}, Turing.jl~\cite{Ge2018}, and the ensemble MCMC sampler emcee~\cite{ForemanMackey2013}. While Bayesian methods provide a formal probabilistic interpretation of parameter uncertainty, they require specification of prior distributions and a scalar likelihood, scale poorly with the number of objectives, and can be expensive for stiff ODE models. On the other hand, multiobjective ensemble methods such as POETs offer a lightweight alternative that naturally handles multiple competing objectives and returns a cloud of near-optimal solutions without requiring probabilistic modeling assumptions.

In this study, we present ParetoEnsembles.jl, a substantially rewritten Julia implementation of the POETs algorithm~\cite{Song2010,Bassen2016}. The package takes a user-defined model and a set of objective functions, each measuring how well a candidate parameter vector reproduces a particular dataset, and returns a collection of parameter sets spanning the trade-off surface, along with their objective values and dominance ranks. Several algorithmic corrections and improvements have been made relative to the original implementation: the dominance check has been corrected from weak to strict Pareto dominance, the per-iteration ranking cost has been reduced from $O(n^{2}m)$ to $O(nm)$ through an incremental update scheme (where $n$ is the archive size and $m$ is the number of objectives), and multi-chain parallel execution has been added to improve front coverage. We demonstrate the package on two biological systems of increasing complexity (\cref{sec:results}), validate it against standard benchmarks and NSGA-II, and assess hyperparameter sensitivity (\cref{sec:si_benchmarks,sec:si_nsga,sec:si_sensitivity}). By making ensemble generation lightweight and accessible, the package lowers the barrier to routine uncertainty characterization in mechanistic modeling.

\section{Methods}\label{sec:methods}

\subsection{Pareto dominance and ranking}\label{sec:dominance}
Consider an archive of $n$ solutions evaluated on $m$ objectives, where lower values are preferred.
Solution $\mathbf{x}_j$ \emph{strictly dominates} solution $\mathbf{x}_i$, written $\mathbf{x}_j \prec \mathbf{x}_i$, if and only if $f_k(\mathbf{x}_j) \leq f_k(\mathbf{x}_i)$ for all $k \in \{1,\ldots,m\}$ and there exists at least one objective $k$ for which $f_k(\mathbf{x}_j) < f_k(\mathbf{x}_i)$ (\cref{eq:dominance}).
The \emph{Pareto rank} of solution $\mathbf{x}_i$ is the number of archive members that strictly dominate it (\cref{eq:rank}), so that solutions with rank zero lie on the Pareto front and identical solutions, which do not dominate each other under strict dominance, share the same rank.
The original POETs implementation~\cite{Song2010,Bassen2016} used \emph{weak} dominance, checking only that $f_k(\mathbf{x}_j) \leq f_k(\mathbf{x}_i)$ for all $k$ without requiring strict inequality in at least one objective; under weak dominance, identical solutions count as dominating each other, which inflates their ranks and causes excessive pruning of good solutions from the archive.
The current implementation enforces strict Pareto dominance. The pairwise test iterates over objectives with an early-exit condition that terminates as soon as any objective of $j$ is found to be worse than $i$, avoiding unnecessary comparisons (pseudocode in \cref{alg:dominates}).
\begin{equation}\label{eq:dominance}
  f_k(\mathbf{x}_j) \leq f_k(\mathbf{x}_i) \quad \forall\, k \in \{1,\ldots,m\}
  \qquad\text{and}\qquad
  \exists\, k : f_k(\mathbf{x}_j) < f_k(\mathbf{x}_i).
\end{equation}
\begin{equation}\label{eq:rank}
  R(\mathbf{x}_i) = \bigl|\{\, j \neq i \mid \mathbf{x}_j \prec \mathbf{x}_i \,\}\bigr|.
\end{equation}

\subsection{Incremental ranking}\label{sec:incremental}
A central computational concern is the cost of ranking. When a single candidate solution $\mathbf{x}_{\text{new}}$ is appended to an archive of $n$ existing solutions with known ranks $\mathbf{R}$, the na\"ive approach recomputes all pairwise dominance checks at $O(n^2 m)$ cost, where $m$ is the number of objectives. Because only one solution changed, however, we can instead perform a single $O(nm)$ pass over the existing archive: for each existing solution $\mathbf{x}_i$, we test dominance in both directions, incrementing $R_i$ if the new solution dominates $\mathbf{x}_i$ and incrementing the new solution's rank if $\mathbf{x}_i$ dominates it. This incremental procedure reduces the per-iteration cost from $O(n^2 m)$ to $O(nm)$, with a full $O(n'^2 m)$ re-rank performed only after removing high-rank solutions from the archive, where $n'$ is the reduced archive size (pseudocode in \cref{alg:rankinsert}).

\subsection{Pareto simulated annealing}\label{sec:annealing}
The complete Pareto simulated annealing procedure combines the dominance check and incremental ranking into a nested loop over temperature levels and candidate evaluations. The algorithm is seeded with an initial parameter vector $\mathbf{x}_0$, an initial temperature $T = 1$, and an archive containing the single evaluated solution $(\mathbf{x}_0, \mathbf{f}(\mathbf{x}_0))$. At each temperature level, $N_{\text{iter}}$ candidate solutions are generated by perturbing the current best parameter vector through a user-supplied neighbor function, and the rank array is updated incrementally at $O(nm)$ cost per candidate.The acceptance probability, computed from the rank array and the current temperature, determines whether the candidate is retained; upon acceptance, solutions whose rank meets or exceeds the cutoff $R_{\text{cutoff}}$ are removed and a full re-rank is performed on the remaining set. If the archive still exceeds the maximum size $n_{\max}$, only the $n_{\max}$ lowest-rank solutions are retained. Upon rejection, the candidate is immediately removed and the rank array is restored from a saved snapshot, a strategy we refer to as \emph{pop-on-reject}. In the original implementation, rejected candidates remained in the archive, causing unbounded growth of $n$ between accepted moves and progressively more expensive ranking operations; the pop-on-reject strategy eliminates this problem. To improve front coverage, multiple independent SA chains can be run from different starting points and their archives merged, with a final re-rank performed on the combined set. This multi-chain strategy provides near-linear parallel speedup and ensures diverse exploration of the trade-off surface (pseudocode for the main loop and multi-chain wrapper in \cref{alg:main,alg:multichain}; software design details and the public API in \cref{sec:si_software}).

\section{Results}\label{sec:results}
We first validated ParetoEnsembles.jl on two standard multiobjective benchmarks, the constrained Binh--Korn problem~\cite{Binh1997} and the unconstrained Fonseca--Fleming problem~\cite{Fonseca1995}, recovering the known Pareto fronts in both cases (\cref{sec:si_benchmarks}). A head-to-head comparison with NSGA-II~\cite{Deb2002} at matched evaluation budgets (${\sim}110{,}000$) showed that ParetoEnsembles.jl achieved comparable hypervolume (the volume of objective space covered by the front, where larger is better), 5--7$\times$ lower inverted generational distance (the average distance from the true front to the computed front, where lower indicates a closer approximation), and 10--30$\times$ denser fronts with thousands of retained solutions compared with ${\sim}200$ for NSGA-II (\cref{sec:si_nsga}).
A hyperparameter sensitivity analysis confirmed robustness to the choice of rank cutoff, cooling rate, and iterations per temperature (\cref{sec:si_sensitivity}). Thus, ParetoEnsembles.jl produced denser and more complete approximations of the trade-off surface, but at the cost of longer wall-clock times (6--150$\times$ slower than NSGA-II, depending on problem size) due to the inherently sequential nature of simulated annealing chains. Next, we focus on two biological applications that illustrate the practical value of ensemble-based parameter estimation.

\subsection{Cell-free gene expression}\label{sec:cellfree}
\begin{figure}[p]
\centering
\includegraphics[width=\textwidth]{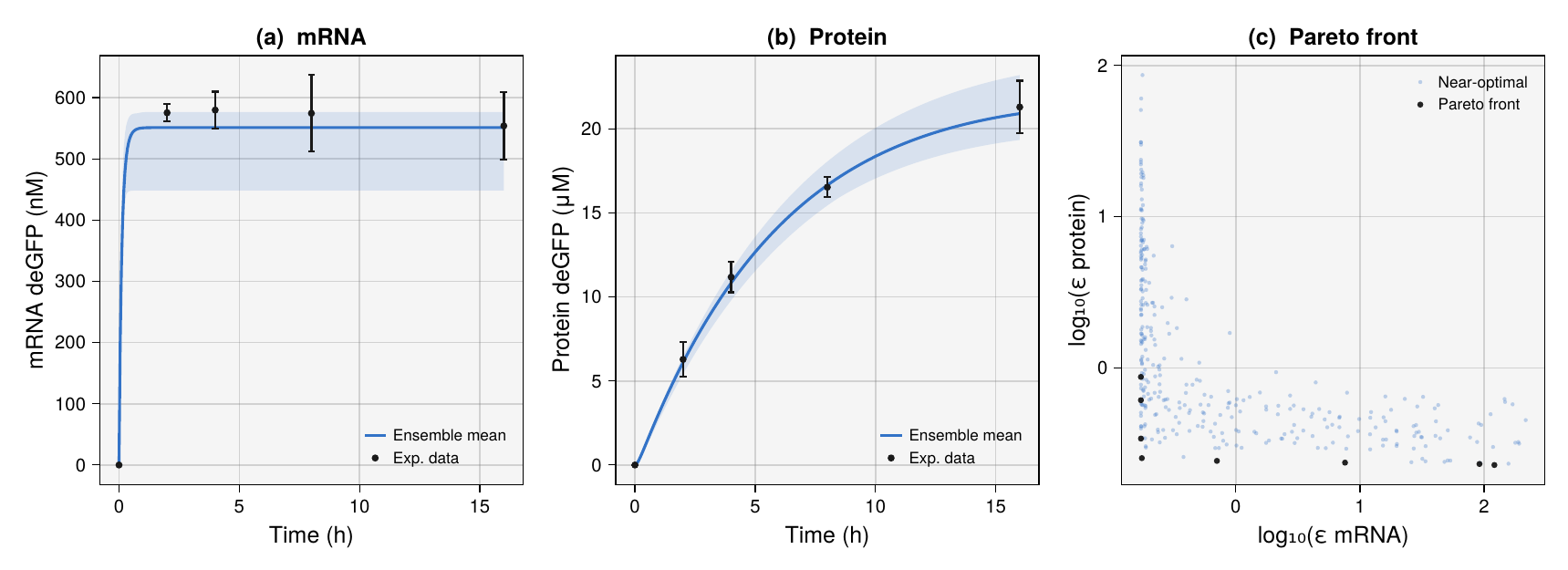}
\caption{Ensemble estimation for a cell-free gene expression circuit ($\sigma_{70} \to \text{P70} \to \text{deGFP}$) fitted to experimental data from~Adhikari et al.~\cite{Adhikari2020}.
(a)~mRNA and (b)~protein concentration versus time; the blue dashed line is the ensemble mean, the shaded region is the 95\% confidence interval, and black points are experimental data with error bars.
(c)~Pareto front showing the trade-off between mRNA and protein fitting error.}\label{fig:cellfree}
\end{figure}

As a first biological application, we considered the estimation of kinetic parameters for a cell-free gene expression circuit. The circuit describes the production of deGFP, a highly translatable variant of enhanced green fluorescent protein developed by Shin and Noireaux~\cite{Shin2010}, driven by the endogenous sigma factor $\sigma_{70}$ via the P70 promoter. We followed the effective biophysical framework of Adhikari et al.~\cite{Adhikari2020} (\cref{fig:cellfree}). The system is governed by two ordinary differential equations for mRNA ($m$) and protein ($p$), where $\dot{m} = \alpha\, u(\sigma_{70}) - \delta_m\, m$ and $\dot{p} = \kappa\, m\, w(t) - \delta_p\, p$, with $u(\sigma_{70}) = \sigma_{70}^{n}/(K^n + \sigma_{70}^n)$ representing a Hill-type promoter activity function and $w(t) = \exp(-\ln 2 \cdot t / \tau_{1/2})$ capturing the experimentally observed decay of translational capacity over the course of the cell-free reaction. Five parameters were treated as unknown (the maximum transcription rate $\alpha$, the effective translation rate constant $\kappa$, the mRNA degradation rate $\delta_m$, the promoter dissociation constant $K$, and the translation capacity half-life $\tau_{1/2}$) while the remaining parameters ($\sigma_{70} = 35$~nM, $n = 1.5$, $\delta_p = 0.005$~h$^{-1}$) were fixed at values taken from the literature.

We fitted the model to experimental data from~Adhikari et al.~\cite{Adhikari2020}, consisting of mRNA and deGFP protein concentrations measured at five time points (0, 2, 4, 8, and 16~h) in triplicate, with reported means and standard deviations. Two objectives were defined as standard-deviation-weighted sums of squared errors: $\varepsilon_{\text{mRNA}} = \sum_i [(m_i^{\text{sim}} - m_i^{\text{obs}}) / \sigma_i^m]^2$ and $\varepsilon_{\text{protein}} = \sum_i [(p_i^{\text{sim}} - p_i^{\text{obs}}) / \sigma_i^p]^2$, and we ran ten parallel chains with a rank cutoff of $R_{\text{cutoff}} = 8$, $N_{\text{iter}} = 50$ candidate solutions per temperature, and a cooling rate of $\alpha = 0.90$, starting from random initial parameter vectors drawn uniformly within the parameter bounds specified in the original study~\cite{Adhikari2020}. The resulting ensemble captured the dynamics of both mRNA and protein expression (\cref{fig:cellfree}a,b): the 95\% confidence interval of the ensemble simulations bracketed the experimental data at all time points, and the ensemble mean tracked the observed trajectories. The Pareto front revealed a clear trade-off between fitting the mRNA and protein time courses (\cref{fig:cellfree}c), with parameter sets that minimized $\varepsilon_{\text{mRNA}}$ tending to have larger $\varepsilon_{\text{protein}}$ and vice versa. This trade-off arises naturally because the mRNA degradation rate and translation capacity half-life have opposing effects on the two species. The ensemble provided a family of plausible parameter sets that traded off between these competing objectives, rather than a single point estimate that favors one objective over the other. Even for this small system, the ensemble revealed trade-off structure among the estimated parameters; we next ask whether the approach scales to a substantially larger and more clinically relevant model.

\subsection{Blood coagulation cascade}\label{sec:coag}
\begin{figure}[p]
\centering
\includegraphics[width=\textwidth]{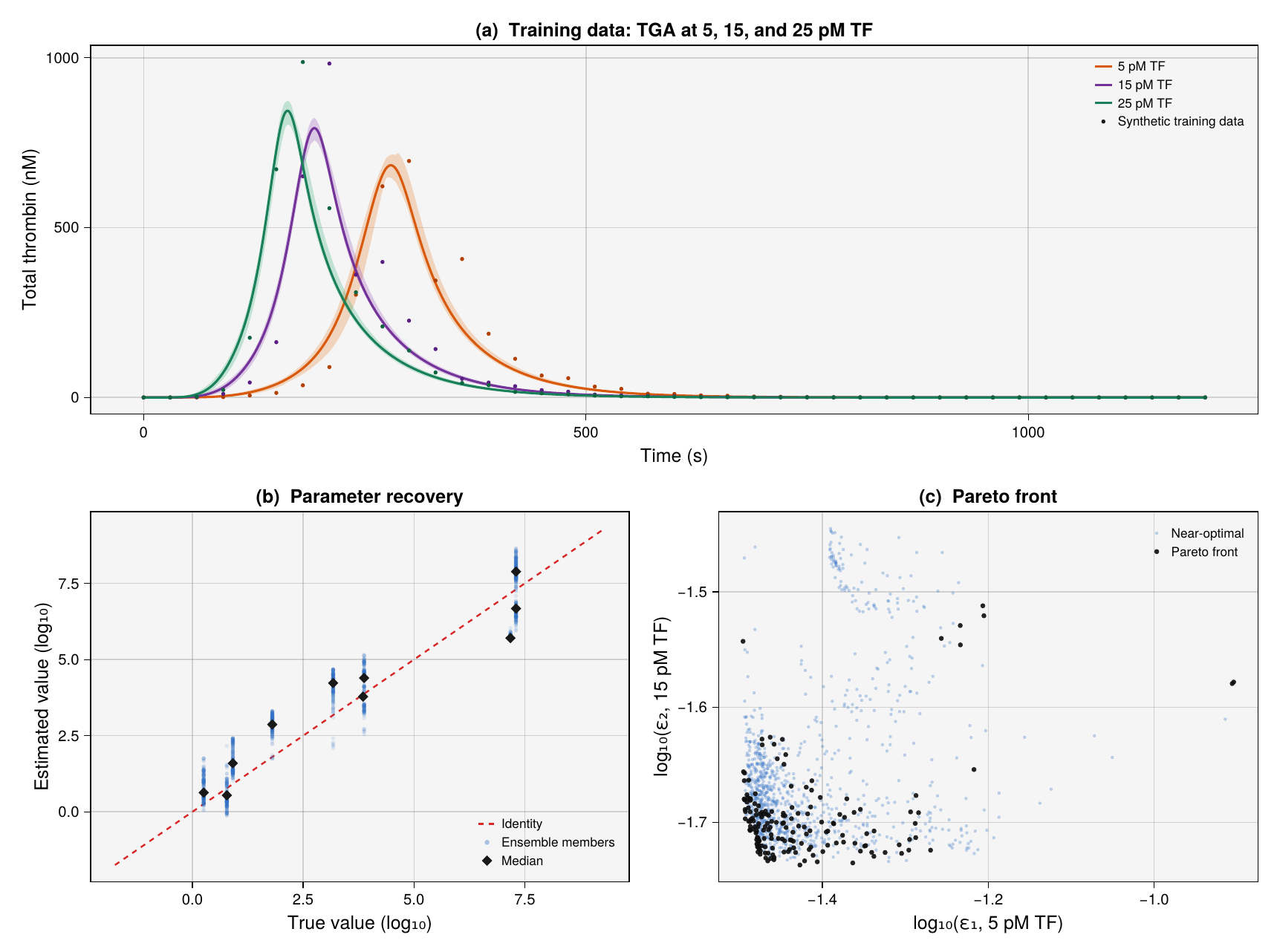}
\caption{Training results for the Hockin--Mann blood coagulation model (34 species, 10 estimated rate constants, 3 data-driven objectives, no regularization).
(a)~Total thrombin concentration versus time at 5~pM (amber), 15~pM (purple), and 25~pM (teal) tissue factor; shaded regions are the ensemble 95\% CI, dashed lines are the ensemble mean, and points are noisy synthetic data (15\% CV).
(b)~Parameter recovery in log-space; the dashed red line is the identity and diamonds show the median ensemble estimate; scatter around each parameter reveals the degree of identifiability from thrombin data alone.
(c)~Pareto front projection ($\varepsilon_1$ vs.\ $\varepsilon_2$, colored by $\varepsilon_3$), with rank-zero solutions in black.}\label{fig:coag}
\end{figure}

To demonstrate the package on a larger-scale system, we considered the Hockin--Mann model of tissue-factor-initiated blood coagulation~\cite{HockinMann2002} (\cref{fig:coag}). This mechanistic ODE model describes 34 chemical species interacting through 27 reactions governed by 42 rate constants, capturing the initiation, amplification, and inhibition phases of thrombin generation. All rate constants are available in the literature and were used as ground truth. We selected ten key catalytic and inhibition rate constants for estimation spanning the extrinsic factor Xase ($k_{\text{cat}}$), intrinsic factor Xase ($k_{\text{cat}}$), prothrombinase ($k_{\text{cat}}$), and antithrombin~III inhibition pathways.
Because these rate constants span six orders of magnitude (from $\sim$1 to $\sim 10^7$), we worked in log-space with bounds of $\pm 1.5$ orders of magnitude around the literature values. We generated noisy synthetic thrombin generation assay (TGA) data by simulating the model with the true rate constants at three tissue factor concentrations (5, 15, and 25~pM) and adding multiplicative Gaussian noise ($x' = x \cdot (1 + \epsilon)$, $\epsilon \sim \mathcal{N}(0, 0.15^2)$) to the total thrombin trajectory at each condition. Three objectives were defined, the normalized sum-of-squared error of the thrombin trajectory at each of the three TF concentrations, with no regularization terms; the parameter recovery reflects only what the training data can constrain rather than being aided by a penalty that pulls estimates toward the known truth.

We ran eight parallel chains with a rank cutoff of $R_{\text{cutoff}} = 10$, $N_{\text{iter}} = 40$ candidate solutions per temperature, and a cooling rate of $\alpha = 0.92$. The resulting ensemble of 222 parameter sets (rank $\leq 1$) produced thrombin trajectories that closely matched the noisy training data at all three TF concentrations (\cref{fig:coag}a), with 95\% confidence intervals that bracketed the data throughout the time course. However, parameter recovery was mixed, revealing the practical identifiability structure of the system (\cref{fig:coag}b). Some rate constants were well recovered (the IIa$+$ATIII inhibition rate to within 2\% and the Factor~Va activation rate to within 5\%), while others were poorly constrained by thrombin data alone: the intrinsic factor Xase $k_{\text{cat}}$ was off by a factor of five and the direct Xa-mediated thrombin generation rate ($k_{\text{Xa} \to \text{IIa}}$) was overestimated by more than twofold. The three-dimensional Pareto front (\cref{fig:coag}c) revealed the expected trade-offs among fitting the three TF conditions, demonstrating that ParetoEnsembles.jl can handle systems of moderate complexity with tens of species, ten or more estimated parameters, and three competing objectives. Several parameters were poorly recovered, yet the model still fit the training data well: this is consistent with compensatory adjustments among rate constants producing nearly identical thrombin trajectories from very different parameter vectors. This motivates the ensemble-based downstream study described next.

\subsection{Ensemble-based prediction and identifiability analysis}\label{sec:ensemble_insights}
\begin{figure}[p]
\centering
\includegraphics[width=\textwidth]{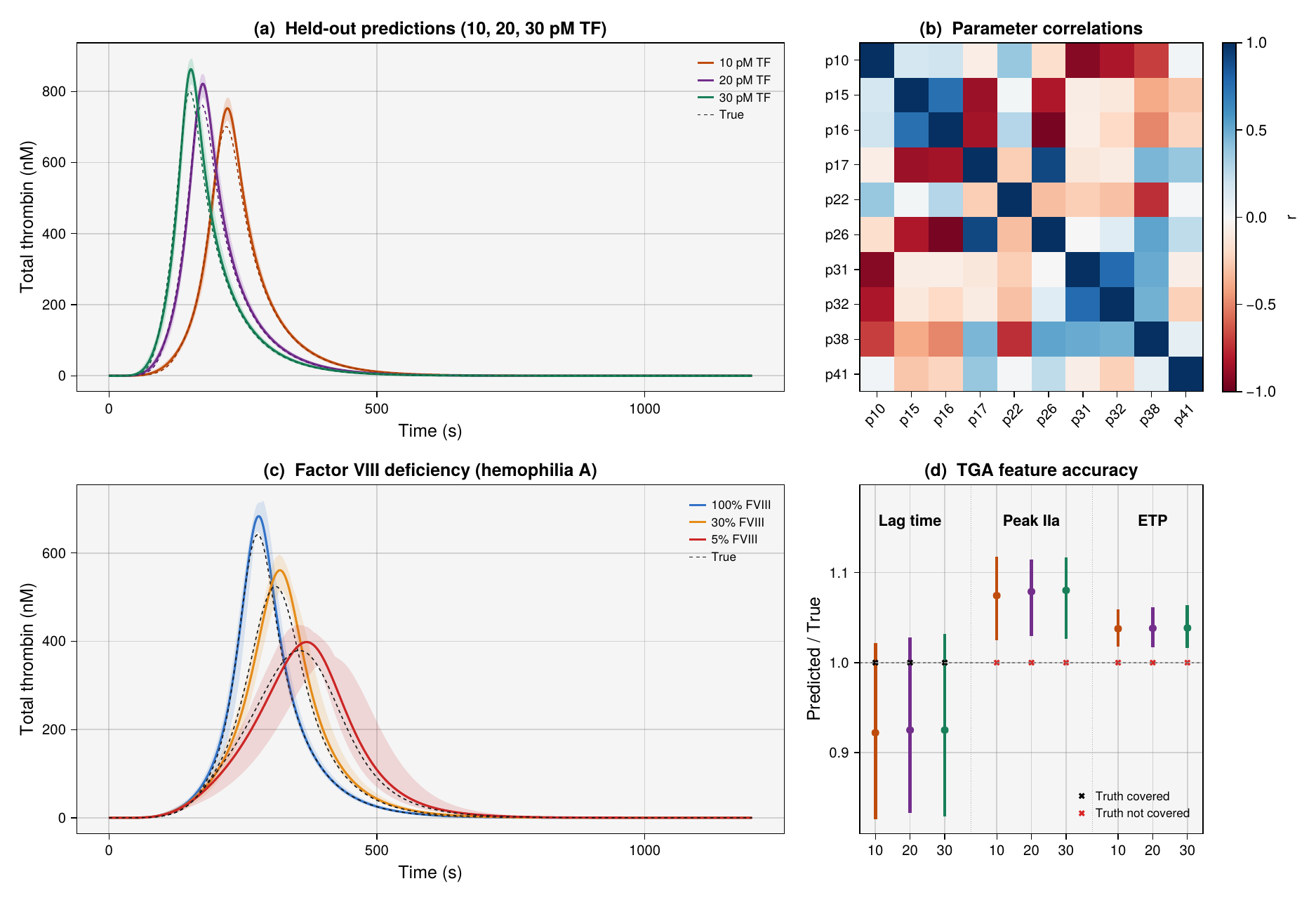}
\caption{Ensemble-based uncertainty characterization for the coagulation model.
(a)~Held-out thrombin predictions at 10, 20, and 30~pM TF (not used during training); solid lines are true trajectories, dashed lines are ensemble means, and shaded regions are 95\% CIs.
The ensemble captures the overall shape but shows a systematic positive bias at peak thrombin.
(b)~Pairwise parameter correlation heatmap revealing extensive compensatory structure; strong negative correlations (e.g., extrinsic Xase $k_{\text{cat}}$ vs.\ prothrombinase $k_{\text{cat}}$, $r = -0.81$; Xa$\to$IIa vs.\ IIa$\to$Va, $r = -0.93$) explain why individual parameters can be poorly recovered while model predictions remain accurate.
(c)~Patient-specific predictions for Factor~VIII deficiency (hemophilia~A) at 100\%, 30\%, and 5\% of nominal FVIII levels; ensemble prediction bands capture the dose-dependent reduction in peak thrombin, and true trajectories (black dashed) fall within all three bands.
(d)~TGA feature accuracy at held-out conditions: ensemble-predicted lag time, peak thrombin, and ETP normalized to true values, with 95\% CIs; black crosses indicate the true value is covered by the interval, red crosses indicate it is not.}\label{fig:insights}
\end{figure}

A key advantage of generating parameter ensembles rather than single point estimates is the ability to propagate uncertainty forward through the model to obtain prediction intervals whose width reflects what the training data can and cannot constrain (\cref{fig:insights}). To test this, we used the coagulation ensemble, which was trained only on TGA curves at 5, 15, and 25~pM TF, to predict the full thrombin time course at three held-out tissue factor concentrations (10, 20, and 30~pM) that were never used during optimization. The ensemble mean tracked the true trajectories at all three conditions, predicting peak thrombin to within 6--7\% of the true values, and the 95\% prediction intervals captured the overall shape of the thrombin generation curve (\cref{fig:insights}a). However, the prediction bands did not fully bracket the true trajectories at the peak, revealing a systematic positive bias in peak amplitude that the ensemble's uncertainty did not account for. The ensemble made this systematic bias visible by providing prediction intervals whose width varied across outputs and conditions. To quantify this further, we extracted three clinically relevant TGA features (lag time, peak thrombin concentration, and endogenous thrombin potential (ETP)) from both the ensemble predictions and the true trajectories at each held-out condition (\cref{fig:insights}d). Lag time was well predicted with appropriately wide 95\% intervals that covered the true value at all three concentrations, while peak thrombin and ETP showed a consistent 4--7\% overprediction with intervals too tight to achieve coverage, indicating that the ensemble is overconfident about thrombin amplitude even as it correctly captures the timing of the coagulation response.

Pairwise correlations in the ensemble revealed the model's identifiability structure (\cref{fig:insights}b). The strongest negative correlation was between the extrinsic factor Xase $k_{\text{cat}}$ and the prothrombinase $k_{\text{cat}}$ ($r = -0.81$), indicating compensatory behavior: when the initiation-phase enzyme is faster, the amplification-phase enzyme must be slower to produce the same thrombin trajectory. Similar compensatory couplings appeared between the direct Xa-mediated thrombin generation rate and Factor~Va activation ($r = -0.93$), reflecting competition between direct and prothrombinase-mediated pathways, and between the TF=VIIa$\to$IX catalytic rate and the mIIa$\to$IIa conversion rate ($r = -0.73$), linking initiation and propagation. These relationships explain why individual parameters can be off by several-fold while model predictions remain accurate to within a few percent: correlated shifts in one rate constant are offset by compensatory changes in others, so the ensemble spans a region of parameter space that produces similar trajectories despite wide variation in individual parameters.

Finally, we used the ensemble to simulate a clinically relevant perturbation: Factor~VIII deficiency (hemophilia~A) at varying severity levels. By reducing the initial Factor~VIII concentration to 30\% (mild hemophilia) and 5\% (severe hemophilia) of the nominal plasma level while keeping all estimated rate constants at their ensemble values, we generated virtual-patient predictions of thrombin generation that captured the expected dose-dependent reduction in peak thrombin and associated uncertainty (\cref{fig:insights}c). Peak thrombin decreased from $684 \pm 22$~nM in normal plasma to $563 \pm 22$~nM for mild hemophilia and $401 \pm 21$~nM for severe hemophilia, with the true trajectories falling within the ensemble prediction bands for all three conditions. The prediction intervals widened as Factor~VIII was reduced, consistent with the expectation that the model becomes less constrained as the system is driven further from the conditions used for training. The ensemble correctly bracketed the true trajectories for these out-of-distribution perturbations despite poor recovery of several individual rate constants, suggesting that ensemble-based uncertainty characterization may be useful in translational settings where model predictions inform clinical decisions. Taken together, these analyses showed that the ensemble revealed identifiability structure, propagated uncertainty to held-out conditions, and supported clinically relevant perturbation studies; but all relied on synthetic data where the model was correct by construction. To test robustness to model error, we repeated the coagulation study with six fixed rate constants perturbed by $\pm 30\%$; the ensemble still predicted held-out peak thrombin to within 4--6\%, though trajectory-level coverage degraded to 31--48\% (\cref{sec:si_misspec}). We next test whether these findings hold when the model is only an approximation of reality.

\subsection{Validation against experimental data}\label{sec:realdata}
\begin{figure}[p]
\centering
\includegraphics[width=\textwidth]{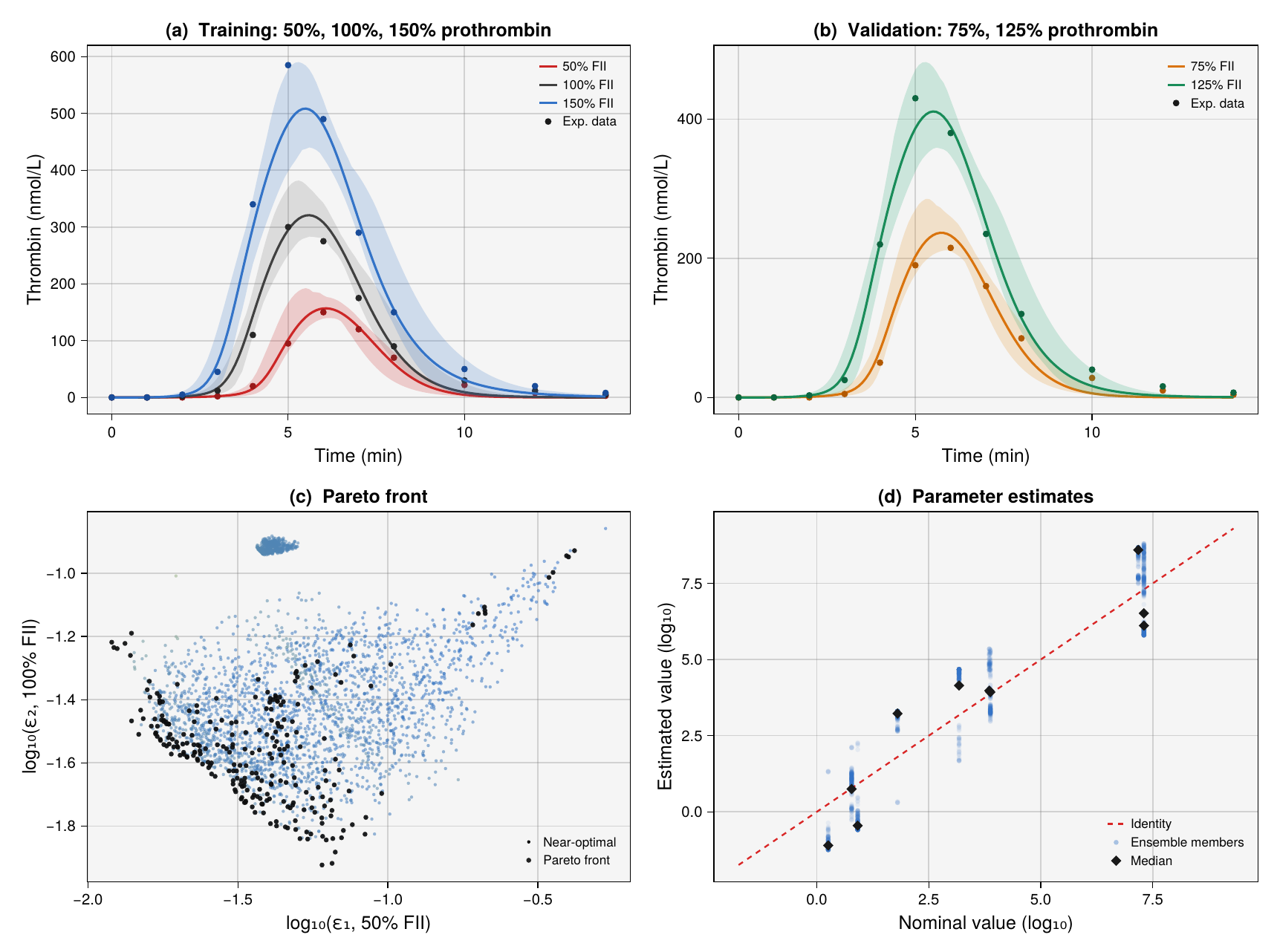}
\caption{Ensemble estimation fitted to experimental thrombin generation data from~Butenas et al.~\cite{Butenas1999}.
Prothrombin was varied from 50\% to 150\% of its mean plasma concentration in a reconstituted synthetic plasma system initiated by 5~pmol/L TF--VIIa.
(a)~Training fits at 50\% (red), 100\% (gray), and 150\% (blue) prothrombin; the ensemble captures the experimental data at low and normal prothrombin levels but underestimates the peak at 150\%, revealing a model--data tension.
(b)~Held-out validation at 75\% (amber) and 125\% (teal) prothrombin; the ensemble predicts peak thrombin to within 1--10\% at conditions never used during training.
(c)~Pareto front projection showing trade-offs among the three training objectives.
(d)~Parameter estimates versus nominal literature values; deviations reflect compensatory adjustments to fit experimental data with an approximate model.}\label{fig:realdata}
\end{figure}

The coagulation analyses above used synthetic data where the ground truth was known, enabling systematic assessment of parameter recovery and coverage. However, these tests left open the question of whether the ensemble approach works when the model is only an approximation of reality. To address this, we fitted the Hockin--Mann model to published experimental data from Butenas et al.~\cite{Butenas1999}, who measured thrombin generation in a reconstituted synthetic plasma system at five prothrombin concentrations ranging from 50\% to 150\% of the mean plasma value, with all other coagulation factors at their nominal concentrations; time-course data were digitized from Figure~3 of~\cite{Butenas1999}, with peak values calibrated against Table~2. This system provided a suitable test case because the initial conditions were well characterized (Table~1 of~\cite{Butenas1999}), the protein C pathway was absent (matching the Hockin--Mann model), and the prothrombin range spans a clinically relevant source of interindividual variability (\cref{fig:realdata}).

We defined three training objectives as the normalized sum of squared errors (SSE) between simulated and experimental thrombin time courses at 50\%, 100\%, and 150\% of normal prothrombin. We held out the 75\% and 125\% conditions for validation. Eight parallel chains were run with the same hyperparameters as the synthetic study, yielding an ensemble of 390 parameter sets (rank $\leq 1$). The ensemble captured the experimental thrombin profiles at the training conditions (\cref{fig:realdata}a), with peak thrombin predicted to within 0.1--10\% across the three prothrombin levels; the residual discrepancies reflected genuine model--data tension in which the Hockin--Mann model's sensitivity to prothrombin concentration did not match the experimental system, a mismatch that was visible in the trade-off structure of the Pareto front (\cref{fig:realdata}c). At the held-out validation conditions, the ensemble predicted peak thrombin at 75\% and 125\% prothrombin to within 1--10\% of the experimental values (\cref{fig:realdata}b), with the ensemble mean tracking the data closely and the 95\% prediction intervals bracketing the experimental time points at both conditions. The estimated rate constants deviated from their nominal literature values (\cref{fig:realdata}d), as expected when fitting real data where the model structure is only an approximation of the underlying biochemistry; these deviations reflected compensatory adjustments in the estimated parameters to offset differences between the model and the experimental system. The spread of the ensemble around each parameter indicated how tightly the data constrained that rate constant. These results confirmed that the ensemble approach generalizes beyond the synthetic setting: even when the model is an imperfect representation of the underlying system, the method produced accurate held-out predictions with informative uncertainty bands.

\section{Discussion}\label{sec:discussion}
We presented ParetoEnsembles.jl, a lightweight Julia implementation of Pareto Optimal Ensemble Techniques (POETs) for multiobjective optimization that addressed several limitations of the original implementation~\cite{Song2010,Bassen2016}. The current version corrected the dominance relation from weak to strict, reduced the per-iteration ranking cost from $O(n^2 m)$ to $O(nm)$ through incremental updating, bounded archive growth through the pop-on-reject strategy and a hard size cap, added both threaded ranking and multi-chain parallel execution, and provided built-in hypervolume computation and convergence diagnostics. The callback-based design decoupled the optimization algorithm from the problem definition, allowing users to apply ParetoEnsembles.jl to any domain where objective functions can be evaluated pointwise. On standard benchmark problems, ParetoEnsembles.jl produced fronts with hypervolume and inverted generational distance competitive with NSGA-II at matched evaluation budgets, while retaining ensembles that were 10--30$\times$ denser than those returned by NSGA-II (\cref{sec:si_nsga}). This density mattered in practice: meaningful computation of prediction intervals, parameter correlations, and coverage statistics from an ensemble required hundreds to thousands of members that fill the near-optimal region of parameter space, not just the 100--200 non-dominated solutions typical of a population-based method. While NSGA-II was considerably faster in wall time, reflecting the inherently sequential nature of SA chains versus the population-parallel evaluation of evolutionary methods, ParetoEnsembles.jl targeted problems where the goal was to characterize the \emph{cloud} of near-optimal parameter sets, not just the front itself. Further, the simplicity of the interface lowered the barrier to generating parameter ensembles as part of a modeling workflow.

The biological applications presented here highlight why ensemble-based parameter estimation differs from conventional single-point optimization. The cell-free gene expression example demonstrated that even a small model with five parameters and two objectives revealed meaningful trade-offs when fitted to real experimental data, and the coagulation cascade example showed that the approach scaled to larger systems with ten parameters spanning six orders of magnitude and three competing objectives. A central finding of the coagulation study was that accurate model predictions do not require accurate parameter recovery. The ensemble predicted thrombin generation at held-out TF concentrations to within 6--7\% of the true peak even though individual rate constants were off by as much as fivefold, because pairwise parameter correlations revealed extensive compensatory structure. The pairwise parameter correlations revealed that many different parameter combinations produced nearly identical thrombin trajectories. The held-out validation also exposed a limitation: while the ensemble correctly predicted the timing of thrombin generation (lag time covered at all conditions), it showed a systematic positive bias in peak amplitude and ETP with confidence intervals too tight to achieve nominal coverage, suggesting that additional data types would be needed to fully constrain the amplitude of the thrombin burst. This controlled experiment validated the algorithm's ability to explore the trade-off surface and provided a ground-truth benchmark for assessing coverage and identifiability, but it did not test robustness to model misspecification. To address this, we repeated the study with six fixed rate constants perturbed by $\pm 30\%$ so that the fitting model was misspecified (\cref{sec:si_misspec}); the ensemble still predicted held-out peak thrombin to within 4--6\%, though trajectory-level coverage degraded to 31--48\%, confirming that the approach is robust to moderate structural error while revealing that parametric uncertainty alone cannot account for model inadequacy. The hemophilia~A simulations demonstrated that the ensemble can propagate parametric uncertainty through clinically relevant perturbations, with prediction intervals that widen as the system is driven further from training conditions and that correctly bracket the true trajectories even when the underlying parameters are individually poorly identified. These results provide a template for how ParetoEnsembles.jl can be used in practice. The hyperparameter sensitivity analysis (\cref{sec:si_sensitivity}) further showed that the algorithm was robust to the choice of rank cutoff, cooling rate, and iterations per temperature, so that users can adopt the default settings without extensive tuning, while the convergence trace feature provided a built-in diagnostic for monitoring annealing progress.

The ensemble produced by ParetoEnsembles.jl is not a Bayesian posterior distribution. The ensemble is a collection of parameter vectors that are near-optimal with respect to multiple objectives, selected by simulated annealing with rank-based acceptance; it does not correspond to samples from a probability distribution conditioned on the data, and the 95\% intervals reported throughout this paper are empirical quantiles of the ensemble predictions, not Bayesian credible intervals. The ensemble makes no claim about the relative probability of different parameter vectors, only that they produce comparably good fits, and its coverage properties depend on how thoroughly the simulated annealing chains explore the near-optimal region rather than on convergence to a stationary distribution.Bayesian approaches such as MCMC~\cite{Carpenter2017,Ge2018} or ensemble MCMC~\cite{ForemanMackey2013} provide a principled probabilistic framework for parameter uncertainty, but they require specification of prior distributions and a scalar likelihood function, which is not naturally defined when multiple incommensurable objectives must be balanced simultaneously; they also face well-known challenges with multimodal posteriors and stiff ODE likelihoods that can make convergence slow or unreliable. The Pareto ensemble approach sidesteps these difficulties at the cost of probabilistic interpretability, and users who require calibrated posterior credible intervals should consider Bayesian methods as a complementary analysis. A promising direction for bridging the two frameworks is the Gibbs posterior, where the loss-based objectives used here replace the log-likelihood in a generalized Bayesian update; importance reweighting of the ensemble members against such a posterior could recover calibrated credible intervals while reusing the expensive model evaluations already performed during optimization. To verify that the coverage patterns are systematic rather than artifacts of a single noise realization, we repeated the synthetic coagulation study five times with independent noise seeds and computed feature-level coverage rates across replicates. The results confirmed a clear and reproducible pattern: lag time and time-to-peak achieved 100\% coverage (5/5 replicates, all held-out conditions), while peak thrombin was covered in only 40\% of replicates and ETP in 60\%, with trajectory-level coverage averaging 87--90\% across conditions. The mean held-out peak error was $7.1 \pm 5.0$\% (range 0.9--17.9\%), confirming that the single-replicate results reported above are representative. This systematic pattern, in which timing is well-calibrated but amplitude is overconfident, likely reflects the concentration of the ensemble along the compensatory structure identified by the correlation analysis: the simulated annealing chains efficiently explore directions in parameter space along which objective values change (i.e., along the Pareto front), but they may underexplore directions orthogonal to the front where objective values are nearly constant yet model predictions differ, leading to prediction intervals that are well-calibrated for some outputs but overconfident for others. We also verified that the downstream predictions are insensitive to the choice of which archive members are included in the ensemble: sweeping the rank cutoff from 0 (Pareto front only) through rank $\leq 5$ changed the mean held-out peak error by less than 0.2 percentage points, confirming that the near-optimal solutions surrounding the front carry essentially the same predictive information as the front itself.

Several features were deliberately omitted from the current scope. The package does not implement gradient-based refinement, surrogate modeling, or adaptive cooling schedules, nor does it provide built-in visualization or benchmark problem libraries. Future directions include integration with Julia's automatic differentiation ecosystem (e.g., ForwardDiff.jl or Enzyme.jl) to enable hybrid gradient-free/gradient-based strategies for problems where gradients are available for some but not all objectives, adaptive archive management using hypervolume indicators~\cite{Zitzler2003} in place of simple rank-based pruning for more principled control over the distribution of retained solutions, and a formal connection to Bayesian inference through the Gibbs posterior framework discussed above. Despite these limitations, the current version provides a minimal and accessible tool for generating Pareto-optimal parameter ensembles, and we anticipate that it will be useful for researchers in systems biology and beyond who need to characterize uncertainty in mechanistic models.

\section{Conclusions}\label{sec:conclusions}
In this study we presented ParetoEnsembles.jl, a Julia package for generating Pareto-optimal parameter ensembles using simulated annealing with incremental dominance ranking. The package corrects a longstanding weak dominance issue in the original POETs implementation, reduces the per-iteration ranking cost from quadratic to linear in the archive size, and adds multi-chain parallel execution that provides improved front coverage from diverse starting points. The package was applied to two biological systems of increasing complexity: a cell-free gene expression circuit fitted to experimental mRNA and protein time-course data, and the Hockin--Mann blood coagulation cascade model with 34 species and ten estimated rate constants spanning six orders of magnitude. A controlled synthetic-data study revealed that accurate model predictions do not require accurate parameter recovery, as the ensemble predicted held-out thrombin generation to within 6--7\% despite individual parameters being off by several-fold. Identifiability analysis, patient-specific hemophilia~A simulations, and a model misspecification study (\cref{sec:si_misspec}) collectively demonstrated that ensemble-based uncertainty characterization reflects what the data can and cannot constrain. Validation against published experimental thrombin generation data from~Butenas et al.~\cite{Butenas1999} confirmed that the approach generalizes beyond the synthetic setting: the ensemble trained on three prothrombin levels predicted held-out conditions to within 1--10\%, even though the model is only an approximation of the underlying experimental system. ParetoEnsembles.jl is open source, registered in the Julia General registry, and can be installed with a single command; we anticipate that its minimal interface and dependency-free design will make it a useful tool for researchers who need to generate parameter ensembles for mechanistic models in systems biology and beyond.

\section*{Acknowledgments}
J.D.V. acknowledges support from the National Heart, Lung, and Blood Institute (NHLBI) of the National Institutes of Health (NIH) under awards R33 HL141787 and R01 HL71944.

\section*{Author Contributions}
\begin{sloppypar}
J.D.V.\ conceived the study, developed the software, performed all analyses, and wrote the manuscript.
\end{sloppypar}

\section*{Competing Interests}
The author declares no competing interests.

\section*{Data and Code Availability}
\begin{sloppypar}
ParetoEnsembles.jl is open source under the MIT license.
The source code is available at
\url{https://github.com/varnerlab/ParetoEnsembles.jl},
and documentation is hosted at
\url{https://varnerlab.github.io/ParetoEnsembles.jl/dev/}.
The package is registered in the Julia General registry and can be installed via \texttt{Pkg.add("ParetoEnsembles")}.
Example code reproducing all results in this paper is included in the \texttt{paper/code} directory of the source repository.
\end{sloppypar}

\bibliography{references}

\clearpage

\clearpage
\appendix
\renewcommand{\thesection}{S\arabic{section}}
\renewcommand{\thetable}{S\arabic{table}}
\renewcommand{\thefigure}{S\arabic{figure}}
\renewcommand{\thealgorithm}{S\arabic{algorithm}}
\setcounter{section}{0}
\setcounter{table}{0}
\setcounter{figure}{0}
\setcounter{algorithm}{0}
\renewcommand{\theHsection}{S\arabic{section}}
\renewcommand{\theHtable}{S\arabic{table}}
\renewcommand{\theHfigure}{S\arabic{figure}}
\providecommand{\theHalgorithm}{\arabic{algorithm}}%
\renewcommand{\theHalgorithm}{S\arabic{algorithm}}

\section*{Supplementary Material}
\addcontentsline{toc}{section}{Supplementary Material}

\section{Algorithm pseudocode}\label{sec:si_algorithms}
This section provides complete pseudocode for the four algorithmic components of ParetoEnsembles.jl.
We begin with the main simulated annealing loop, which orchestrates the optimization, and then present the subroutines it relies on: the strict dominance test that underlies all ranking decisions, the incremental rank update that keeps per-iteration cost linear, and the multi-chain wrapper that parallelizes the search.

\paragraph{Main algorithm.}
The core of ParetoEnsembles.jl is a Pareto simulated annealing loop (\cref{alg:main}) that maintains an archive of evaluated solutions and their Pareto ranks.
At each temperature level, candidates are generated by perturbing the current best parameter vector, and the archive is updated incrementally via \textsc{RankInsert}.
Accepted candidates trigger a pruning step that removes highly dominated solutions (rank $\geq R_{\text{cutoff}}$) and enforces a hard archive size cap, followed by a full re-rank of the smaller remaining set.
Rejected candidates are immediately removed and the rank array is restored from a snapshot, the \emph{pop-on-reject} strategy described in the main text, preventing unbounded archive growth between accepted moves.

\begin{algorithm}[H]
\caption{\textsc{ParetoEnsembles}: Pareto simulated annealing with incremental ranking}\label{alg:main}
\begin{algorithmic}[1]
\Require User callbacks: \textsc{Objective}, \textsc{Neighbor}, \textsc{Accept}, \textsc{Cool}
\Require Initial parameter vector $\mathbf{x}_0$; hyperparameters $N_{\text{iter}}$, $R_{\text{cutoff}}$, $T_{\min}$, $n_{\max}$
\Ensure Archive of solutions with objective values, parameters, and Pareto ranks
\Statex
\State $\mathbf{x}_{\text{best}} \gets \mathbf{x}_0$
\State $\mathcal{A} \gets \{(\mathbf{x}_0,\; \textsc{Objective}(\mathbf{x}_0))\}$; \quad $\mathbf{R} \gets [0]$; \quad $T \gets 1$
\Statex
\While{$T > T_{\min}$}
  \For{$t = 1, \ldots, N_{\text{iter}}$}
    \State $\mathbf{x}' \gets \textsc{Neighbor}(\mathbf{x}_{\text{best}})$; \quad $\mathbf{f}' \gets \textsc{Objective}(\mathbf{x}')$
    \State $\mathbf{R}_{\text{saved}} \gets \text{copy}(\mathbf{R})$
    \State Append $(\mathbf{x}', \mathbf{f}')$ to $\mathcal{A}$; \quad \Call{RankInsert}{$\mathcal{A}, \mathbf{R}$}
    \If{$\textsc{Accept}(\mathbf{R}, T) > U(0,1)$}
      \State Prune: $\mathcal{A} \gets \{(\mathbf{x}_i, \mathbf{f}_i) \in \mathcal{A} \mid R_i < R_{\text{cutoff}}\}$
      \State $\mathbf{R} \gets \textsc{FullRank}(\mathcal{A})$
      \If{$|\mathcal{A}| > n_{\max}$}
        \State Keep only the $n_{\max}$ solutions with lowest $R_i$
        \State $\mathbf{R} \gets \textsc{FullRank}(\mathcal{A})$
      \EndIf
      \State $\mathbf{x}_{\text{best}} \gets \mathbf{x}'$
    \Else
      \State Remove $(\mathbf{x}', \mathbf{f}')$ from $\mathcal{A}$; \quad $\mathbf{R} \gets \mathbf{R}_{\text{saved}}$
    \EndIf
  \EndFor
  \State $T \gets \textsc{Cool}(T)$
\EndWhile
\State $\mathbf{R} \gets \textsc{FullRank}(\mathcal{A})$
\State \Return $\mathcal{A},\, \mathbf{R}$
\end{algorithmic}
\end{algorithm}

\paragraph{Strict dominance test.}
The pairwise dominance check (\cref{alg:dominates}) determines whether one solution strictly dominates another by iterating over all $m$ objectives.
The test returns \textsc{true} only if solution $j$ is no worse than $i$ on every objective and strictly better on at least one, which is the strict Pareto dominance relation that corrects the weak dominance used in the original POETs implementation.
An early-exit condition terminates the loop as soon as $j$ is found to be worse than $i$ on any single objective, avoiding unnecessary comparisons.

\begin{algorithm}[H]
\caption{\textsc{StrictlyDominates}$(\mathbf{f}_j, \mathbf{f}_i, m)$: Does solution $j$ strictly dominate solution $i$?}\label{alg:dominates}
\begin{algorithmic}[1]
\Require Objective vectors $\mathbf{f}_j, \mathbf{f}_i \in \mathbb{R}^m$ (lower is better)
\Ensure \textsc{true} if $\mathbf{f}_j \prec \mathbf{f}_i$, \textsc{false} otherwise
\State $\textit{strictly\_better} \gets \textsc{false}$
\For{$k = 1, \ldots, m$}
  \If{$f_{j,k} > f_{i,k}$}
    \State \Return \textsc{false}
    \Comment{$j$ is worse in objective $k$}
  \ElsIf{$f_{j,k} < f_{i,k}$}
    \State $\textit{strictly\_better} \gets \textsc{true}$
    \Comment{$j$ better in $\geq 1$ objective}
  \EndIf
\EndFor
\State \Return $\textit{strictly\_better}$
\end{algorithmic}
\end{algorithm}

\paragraph{Incremental rank update.}
When a single candidate is appended to the archive, recomputing all pairwise dominance relationships would cost $O(n^2 m)$.
The \textsc{RankInsert} procedure (\cref{alg:rankinsert}) exploits the fact that only one solution changed: it performs a single pass over the existing archive, testing dominance in both directions between the new solution and each incumbent.
Existing ranks are incremented where the new solution dominates an incumbent, and the new solution's rank is accumulated from the number of incumbents that dominate it.
This reduces the per-iteration cost to $O(nm)$, with a full $O(n'^2 m)$ re-rank deferred to the pruning step where the archive is smaller.

\begin{algorithm}[H]
\caption{\textsc{RankInsert}$(\mathcal{A}, \mathbf{R})$: Incremental rank update after appending one solution}\label{alg:rankinsert}
\begin{algorithmic}[1]
\Require Archive $\mathcal{A} = \{(\mathbf{x}_1, \mathbf{f}_1), \ldots, (\mathbf{x}_n, \mathbf{f}_n)\}$ where $(\mathbf{x}_n, \mathbf{f}_n)$ is new
\Require Rank array $\mathbf{R} = [R_1, \ldots, R_{n-1}]$ valid for solutions $1, \ldots, n{-}1$
\Ensure Updated $\mathbf{R} = [R_1, \ldots, R_n]$ valid for the full archive
\State $R_{\text{new}} \gets 0$
\For{$i = 1, \ldots, n-1$}
  \If{\Call{StrictlyDominates}{$\mathbf{f}_n, \mathbf{f}_i, m$}}
    \State $R_i \gets R_i + 1$
    \Comment{new solution dominates $i$}
  \EndIf
  \If{\Call{StrictlyDominates}{$\mathbf{f}_i, \mathbf{f}_n, m$}}
    \State $R_{\text{new}} \gets R_{\text{new}} + 1$
    \Comment{$i$ dominates new solution}
  \EndIf
\EndFor
\State Append $R_{\text{new}}$ to $\mathbf{R}$
\end{algorithmic}
\end{algorithm}

\paragraph{Multi-chain parallel execution.}
Because simulated annealing is inherently sequential within a single chain, the most effective parallelization strategy is to run $C$ independent chains from different starting points and merge the results (\cref{alg:multichain}).
Each chain receives its own deterministic random number generator seeded from a base seed and the chain index, ensuring full reproducibility.
After all chains complete, their archives are concatenated and a final full ranking is performed on the merged set, producing a combined ensemble with broader coverage of the trade-off surface than any individual chain can achieve.

\begin{algorithm}[H]
\caption{\textsc{ParetoEnsemblesParallel}: Multi-chain parallel execution}\label{alg:multichain}
\begin{algorithmic}[1]
\Require Initial states $\{\mathbf{x}_0^{(1)}, \ldots, \mathbf{x}_0^{(C)}\}$; base seed $s$
\Ensure Merged archive with Pareto ranks
\For{$c = 1, \ldots, C$ \textbf{in parallel}}
  \State $\text{rng}_c \gets \textsc{Seed}(s, c)$
  \State $(\mathcal{A}_c, \mathbf{R}_c) \gets \textsc{ParetoEnsembles}(\mathbf{x}_0^{(c)},\, \text{rng}_c)$
\EndFor
\State $\mathcal{A} \gets \mathcal{A}_1 \cup \cdots \cup \mathcal{A}_C$
\State $\mathbf{R} \gets \textsc{FullRank}(\mathcal{A})$
\State \Return $\mathcal{A},\, \mathbf{R}$
\end{algorithmic}
\end{algorithm}

\clearpage
\section{Software design}\label{sec:si_software}
ParetoEnsembles.jl is a pure Julia package with no dependencies beyond the standard library module \texttt{Random}, and its public API consists of five functions (\cref{tab:api}).
The primary entry point, \texttt{estimate\_ensemble}, runs a single Pareto SA chain and returns the retained objective values, parameter vectors, and Pareto ranks; its parallel counterpart, \texttt{estimate\_ensemble\_parallel}, runs $C$ independent chains concurrently, merges their archives, and re-ranks the combined set.
A standalone \texttt{rank\_function} computes Pareto ranks for an arbitrary $m \times n$ objective matrix and can be used independently of the optimization loop, while \texttt{hypervolume} computes the hypervolume indicator for a 2D front using a sweep-line algorithm in $O(n \log n)$ time and \texttt{pareto\_front} extracts the rank-zero subset from the archive for downstream analysis.

\begin{sloppypar}
The algorithm is defined by four user-supplied callback functions (\cref{tab:callbacks}): an objective function that maps a parameter vector $\mathbf{x} \in \mathbb{R}^d$ to an $m \times 1$ array of objective values, a neighbor function that perturbs a given parameter vector to generate a candidate, an acceptance probability function that returns a scalar probability from the rank array and temperature, and a cooling function that maps the current temperature to a reduced value.
Beyond these four callbacks, the behavior of \texttt{estimate\_ensemble} is controlled by keyword arguments including the pruning threshold \texttt{rank\_cutoff} (default~5), the number of candidates per temperature level \texttt{maximum\_number\_of\_iterations} (default~20), the stopping temperature \texttt{temperature\_min} (default $10^{-4}$), and the hard archive size cap \texttt{maximum\_archive\_size} (default~1000).
When \texttt{trace=true}, the algorithm records convergence diagnostics (temperature, archive size, and hypervolume) at each cooling step, returning a fourth element in the output tuple that allows users to monitor convergence and assess whether the annealing schedule was sufficient.
\end{sloppypar}

\begin{table}[ht]
\centering
\caption{Public API of ParetoEnsembles.jl.}\label{tab:api}
\begin{tabular}{@{}lp{7.5cm}@{}}
\toprule
\textbf{Function} & \textbf{Description} \\
\midrule
\texttt{estimate\_ensemble} & Run a single Pareto SA chain. Returns objective values, parameter vectors, and Pareto ranks for the retained archive. \\
\texttt{estimate\_ensemble\_parallel} & Run $C$ independent chains in parallel (one per starting point), merge archives, and re-rank. \\
\texttt{rank\_function} & Compute Pareto ranks for an $m \times n$ objective matrix. \\
\texttt{hypervolume} & Compute the hypervolume indicator for a 2D Pareto front given a reference point. \\
\texttt{pareto\_front} & Extract Pareto-optimal (rank $= 0$) solutions from the archive. \\
\bottomrule
\end{tabular}
\end{table}

\begin{table}[ht]
\centering
\caption{User-supplied callback functions.}\label{tab:callbacks}
\small
\begin{tabular}{@{}lp{2.8cm}p{4.5cm}@{}}
\toprule
\textbf{Callback} & \textbf{Signature} & \textbf{Purpose} \\
\midrule
\texttt{objective\_function} & $(\mathbf{x}) \to \mathbb{R}^{m \times 1}$ & Evaluate $m$ objectives at parameter vector $\mathbf{x}$ \\
\texttt{neighbor\_function} & $(\mathbf{x}) \to \mathbb{R}^{d}$ & Generate a candidate by perturbing $\mathbf{x}$ \\
\texttt{acceptance\_probability\_function} & $(\mathbf{R}, T) \to [0,1]$ & Acceptance probability given ranks and temperature \\
\texttt{cooling\_function} & $(T) \to T'$ & Annealing schedule \\
\bottomrule
\end{tabular}
\end{table}

\section{Parallel execution}\label{sec:si_parallel}
Two opportunities for parallelism arise from the structure of the algorithm.
First, the outer loop of both the full ranking and the incremental \textsc{RankInsert} procedure iterates independently over archive solutions, so setting \texttt{parallel\_evaluation=true} dispatches to threaded variants that distribute these independent iterations across available threads using Julia's \texttt{Threads.@threads} construct, with an atomic accumulator used for $R_{\text{new}}$ in the incremental case since multiple threads contribute to it.
Second, and more significantly, because simulated annealing is inherently sequential within a single chain, the most effective parallelization strategy is to run $C$ independent chains from different starting points and merge the resulting archives (\cref{alg:multichain}); each chain receives its own deterministic random number generator seeded from a base seed and the chain index, ensuring reproducibility, and after all chains complete, the archives are concatenated and a final full ranking is performed on the merged set.

To quantify the benefit of multi-chain parallelism, we measured wall clock time for the cell-free ensemble estimation (ten chains, $N_{\text{iter}} = 50$) as a function of the number of Julia threads (\cref{tab:scaling}).
Each configuration was run five times and the median time is reported; the speedup is approximately linear up to four threads ($3.1\times$) and reaches $4.1\times$ at eight threads, with diminishing returns due to load imbalance when ten chains are distributed across eight threads and the serial cost of the final archive merge and re-ranking step.

\begin{table}[ht]
\centering
\caption{Multi-chain parallel scaling on the cell-free gene expression model (ten chains, $N_{\text{iter}} = 50$, median of five runs).}\label{tab:scaling}
\begin{tabular}{@{}ccc@{}}
\toprule
\textbf{Threads} & \textbf{Wall clock (s)} & \textbf{Speedup} \\
\midrule
1 & 18.52 & 1.0$\times$ \\
2 &  9.38 & 2.0$\times$ \\
4 &  5.94 & 3.1$\times$ \\
8 &  4.51 & 4.1$\times$ \\
\bottomrule
\end{tabular}
\end{table}

\clearpage
\section{Standard benchmarks}\label{sec:si_benchmarks}
\begin{figure}[ht]
\centering
\includegraphics[width=\textwidth]{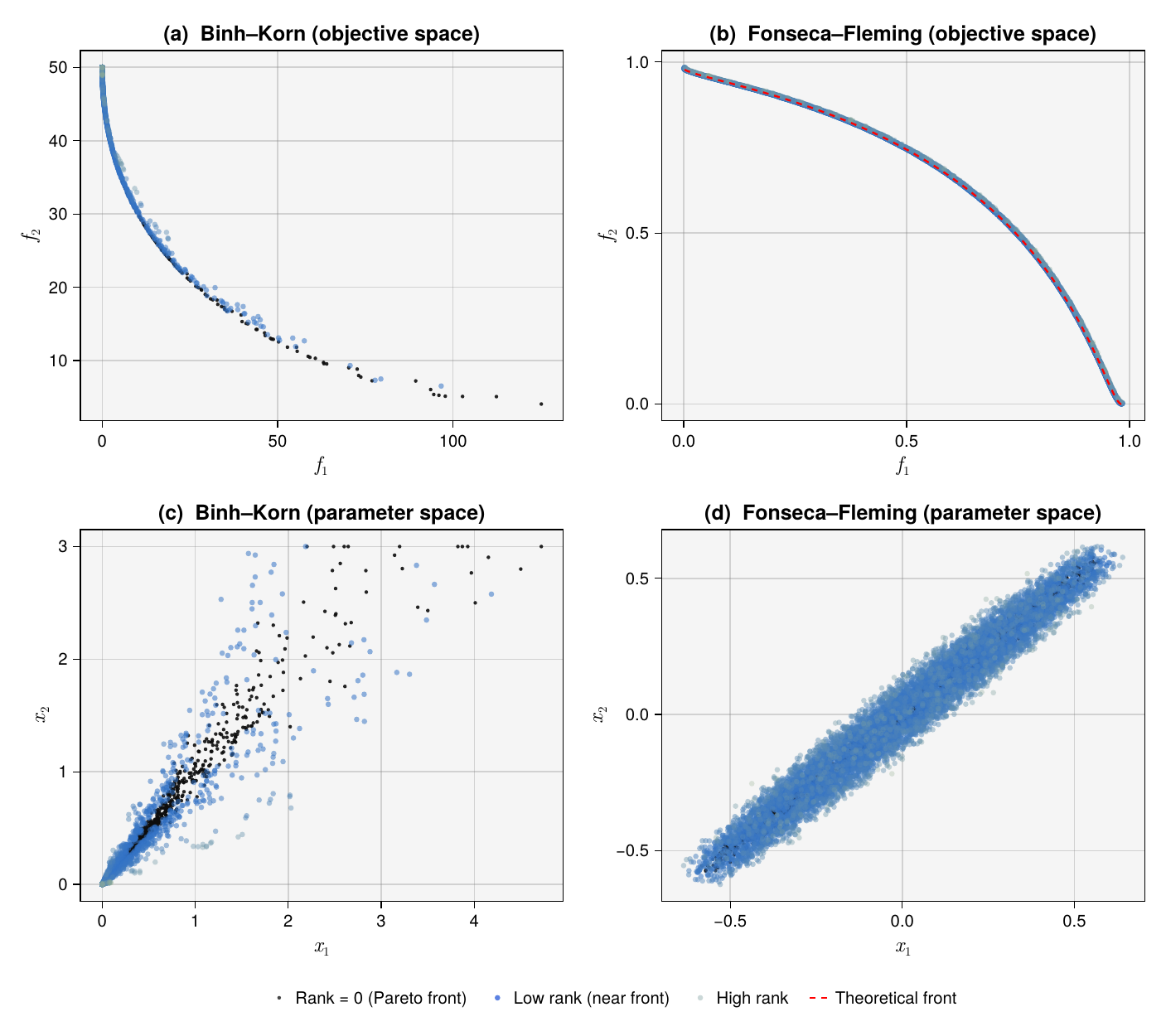}
\caption{Benchmark results for (a,c) Binh--Korn and (b,d) Fonseca--Fleming ($d=3$).
Top row: objective space; bottom row: parameter-space projections ($x_1$ vs.\ $x_2$).
Non-dominated solutions (rank~$= 0$, dark) define the Pareto front, while near-optimal solutions (blue, colored by rank) form a cloud around it representing the retained ensemble.
The dashed red curve in~(b) is the theoretical Pareto front.
Ten parallel chains were used with $R_{\text{cutoff}} = 12$, $N_{\text{iter}} = 60$, and $\alpha = 0.95$.}\label{fig:benchmarks}
\end{figure}

\begin{figure}[ht]
\centering
\includegraphics[width=\textwidth]{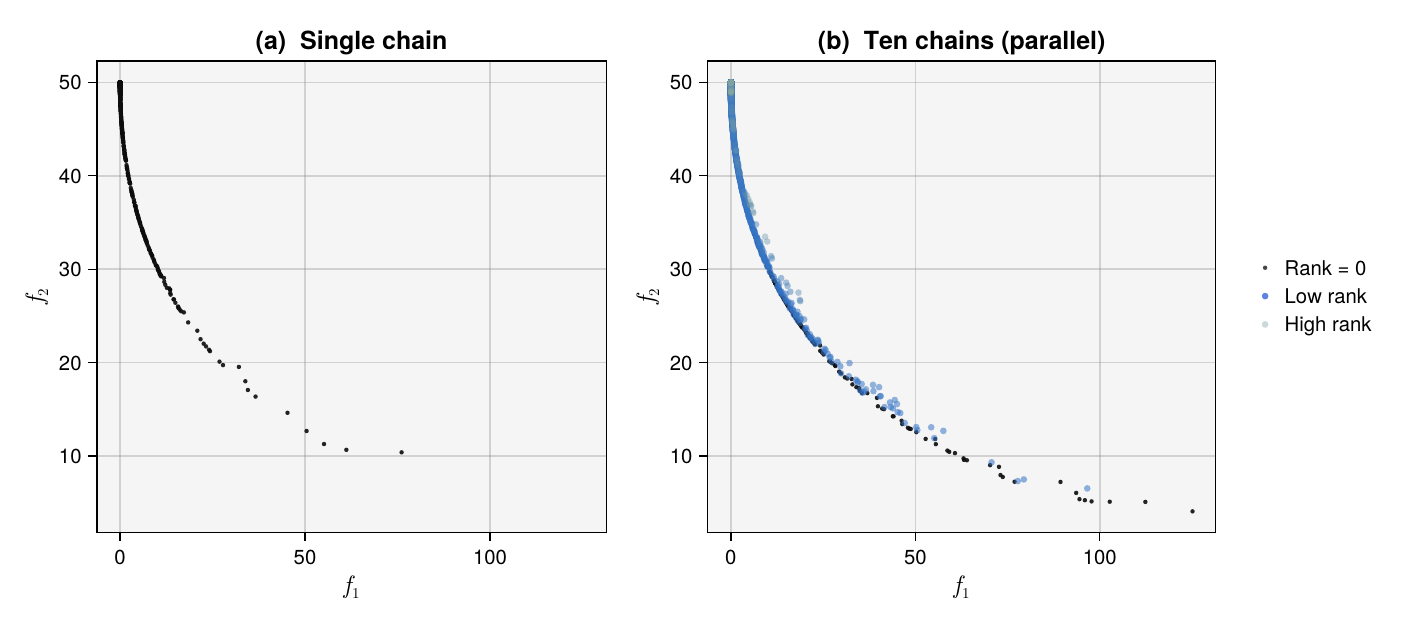}
\caption{Comparison of (a) single-chain and (b) ten-chain parallel execution on the Binh--Korn problem.
Both panels share the same axis limits.
The single chain yields a sparse archive of non-dominated solutions, while the merged multi-chain archive provides denser front coverage and a visible cloud of near-optimal solutions (blue) surrounding the front.}\label{fig:parallel}
\end{figure}

We validated ParetoEnsembles.jl on two standard multiobjective benchmarks: the constrained Binh--Korn problem~\cite{Binh1997} (\cref{eq:bk_obj}) and the unconstrained Fonseca--Fleming problem~\cite{Fonseca1995} with $d = 3$ decision variables (\cref{eq:ff,fig:benchmarks}).
Constraints in the Binh--Korn problem are enforced via quadratic penalty terms added to the objective values, and bound constraints are handled by clamping in the neighbor function.
We ran both benchmarks using ten parallel chains with $N_{\text{iter}} = 60$ candidates per temperature, a rank cutoff of $R_{\text{cutoff}} = 12$, and a cooling rate of $\alpha = 0.95$, and the resulting Pareto fronts in objective space recover the characteristic trade-off curves for both problems, with a visible cloud of near-optimal solutions surrounding the non-dominated front (\cref{fig:benchmarks}a,b).
The computed front for the Fonseca--Fleming case lies directly on the theoretical curve, confirming that the algorithm converges to the correct solution set, and in parameter space the Binh--Korn Pareto-optimal decision vectors trace a curved manifold from the origin toward $(5, 3)$ while the Fonseca--Fleming solutions cluster along the diagonal $x_1 \approx x_2$ as expected from the symmetric structure of the objectives (\cref{fig:benchmarks}c,d).
Comparing a single chain against ten parallel chains on the Binh--Korn problem illustrates the benefit of multi-chain execution (\cref{fig:parallel}): the single chain recovers the trade-off curve but, because every retained solution is non-dominated within that chain's archive, the rank array contains only rank-zero entries, whereas the merged multi-chain run produces a denser front with broader coverage of the feasible region.
\begin{align}
  \min_{\mathbf{x}} \quad & f_1 = 4x_1^2 + 4x_2^2, \qquad f_2 = (x_1 - 5)^2 + (x_2 - 5)^2 \label{eq:bk_obj} \\
  \text{s.t.} \quad & (x_1 - 5)^2 + x_2^2 \leq 25, \qquad (x_1 - 8)^2 + (x_2 - 3)^2 \geq 7.7 \notag \\
  & 0 \leq x_1 \leq 5, \quad 0 \leq x_2 \leq 3. \notag
\end{align}
\begin{equation}\label{eq:ff}
  f_1(\mathbf{x}) = 1 - \exp\!\Biggl(-\sum_{i=1}^{d}\Bigl(x_i - \tfrac{1}{\sqrt{d}}\Bigr)^{\!2}\Biggr), \qquad
  f_2(\mathbf{x}) = 1 - \exp\!\Biggl(-\sum_{i=1}^{d}\Bigl(x_i + \tfrac{1}{\sqrt{d}}\Bigr)^{\!2}\Biggr).
\end{equation}

\clearpage
\section{Comparison with NSGA-II}\label{sec:si_nsga}
\begin{figure}[ht]
\centering
\includegraphics[width=\textwidth]{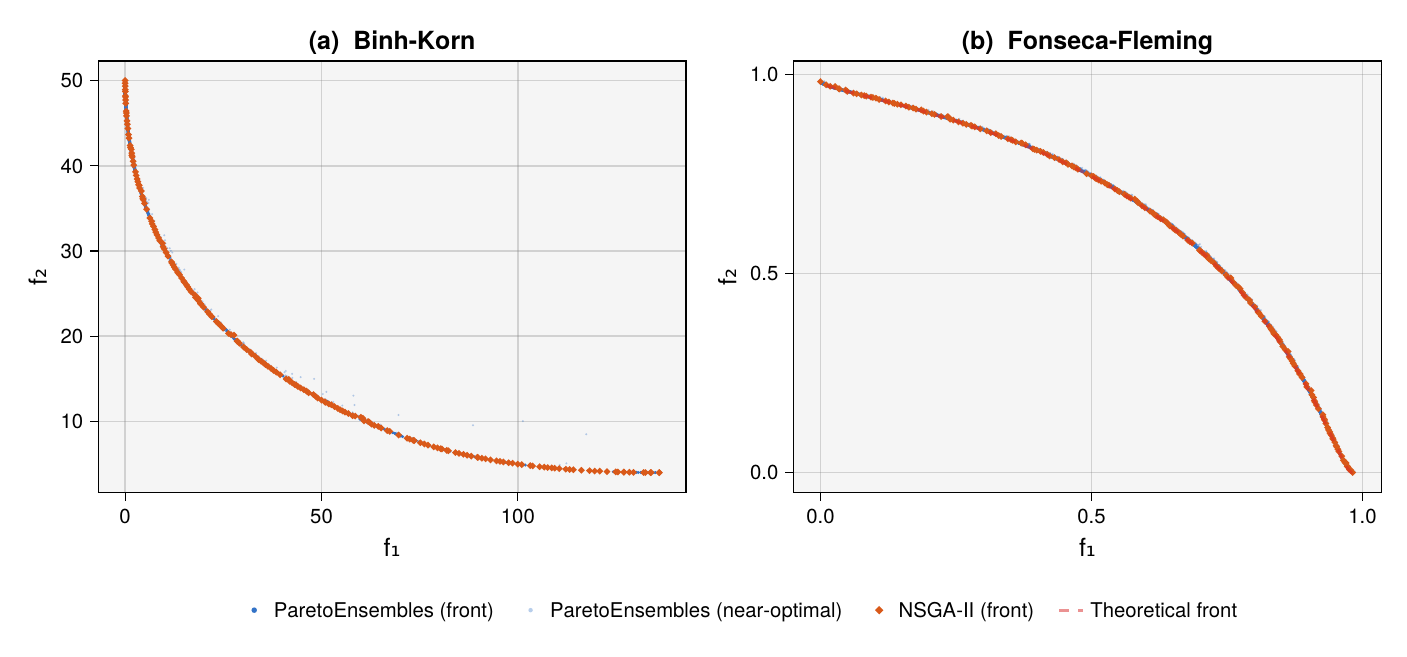}
\caption{Comparison of ParetoEnsembles.jl (left) and NSGA-II (right) on (a,b)~Binh--Korn and (c,d)~Fonseca--Fleming at matched evaluation budgets ($\sim$110{,}000).
ParetoEnsembles.jl retains a cloud of near-optimal solutions (blue) in addition to the Pareto front (black), while NSGA-II returns only the non-dominated front.
The dashed red curve in (c,d) is the theoretical Fonseca--Fleming front.}\label{fig:comparison}
\end{figure}

To position ParetoEnsembles.jl relative to population-based evolutionary methods, we compared it against NSGA-II~\cite{Deb2002} as implemented in Metaheuristics.jl~\cite{Mejia2022} on both benchmark problems, matching the function evaluation budget at approximately 110{,}000 evaluations per solver (ParetoEnsembles: 10 chains, $N_{\text{iter}} = 60$ candidates per temperature, cooling rate $\alpha = 0.95$; NSGA-II: population 200, 550 generations).
Five independent replicates were run for each solver, and the median hypervolume indicator (HV) and inverted generational distance (IGD) are reported in \cref{tab:comparison}, with IGD computed against the analytical Pareto front for the Fonseca--Fleming problem and against a densely sampled reference front from a high-budget run for Binh--Korn.
ParetoEnsembles.jl achieved comparable or higher hypervolume than NSGA-II on both problems and substantially lower IGD (5$\times$ lower on Binh--Korn and 7$\times$ lower on Fonseca--Fleming), indicating a closer approximation to the true Pareto front (\cref{fig:comparison}), while also producing much denser fronts with 2{,}000--6{,}000 non-dominated solutions compared with 200 for NSGA-II, reflecting its design as an ensemble generator that characterizes the full cloud of near-optimal solutions rather than merely approximating the front boundary.
NSGA-II was 6--150$\times$ faster in wall time depending on problem size, consistent with its population-parallel evaluation strategy, but this speed advantage comes at the cost of a sparser representation of the trade-off surface that may be insufficient for downstream uncertainty analysis.

\begin{table}[ht]
\centering
\caption{Comparison of ParetoEnsembles.jl and NSGA-II on benchmark problems (median over 5 replicates). Higher HV and lower IGD are better. Both solvers used $\sim$110{,}000 function evaluations.}\label{tab:comparison}
\begin{tabular}{@{}llcccc@{}}
\toprule
\textbf{Problem} & \textbf{Solver} & \textbf{HV} & \textbf{IGD} & \textbf{Front size} & \textbf{Time (s)} \\
\midrule
Binh--Korn & ParetoEnsembles & \textbf{8035} & \textbf{0.014} & 6172 & 77 \\
           & NSGA-II         & 8027          & 0.071          & 200  & 0.5 \\
\midrule
Fonseca--Fleming & ParetoEnsembles & \textbf{0.44} & \textbf{0.0004} & 2051 & 3.1 \\
                 & NSGA-II         & 0.44          & 0.003           & 200  & 0.5 \\
\bottomrule
\end{tabular}
\end{table}

\clearpage
\section{Hyperparameter sensitivity}\label{sec:si_sensitivity}
\begin{figure}[ht]
\centering
\includegraphics[width=\textwidth]{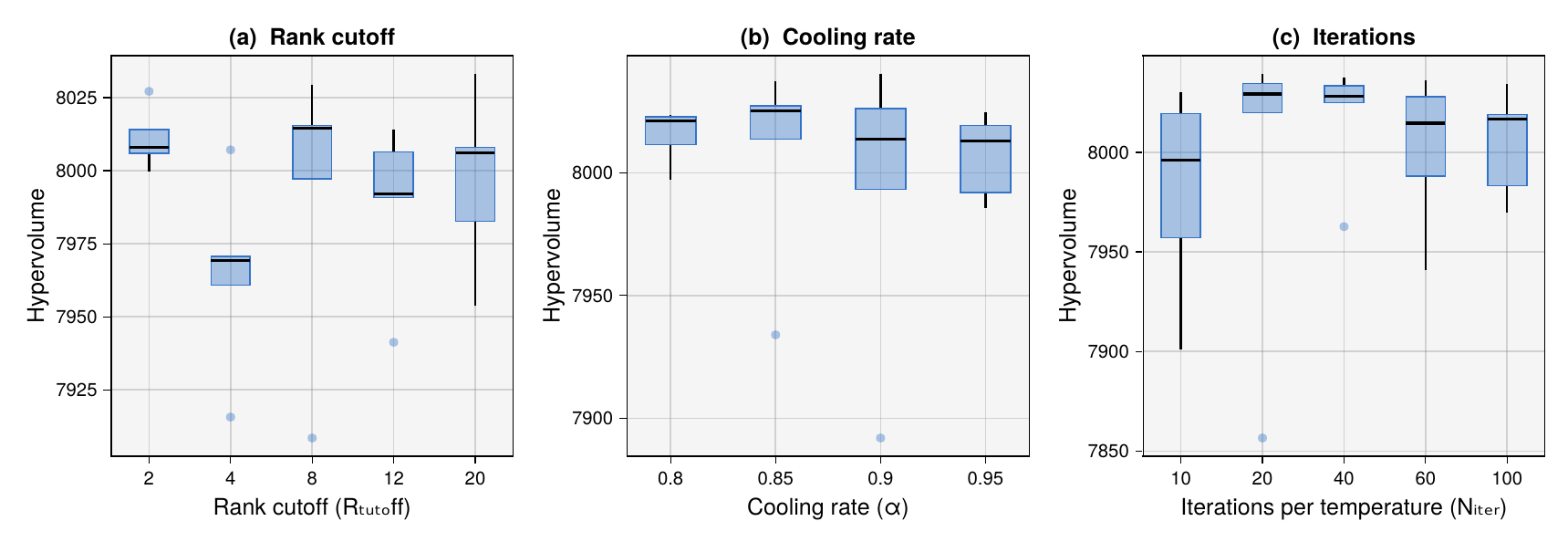}
\caption{Hyperparameter sensitivity on the Binh--Korn benchmark.
Box plots show the distribution of hypervolume over 5 replicates (median, interquartile range, and whiskers).
(a)~Rank cutoff $R_{\text{cutoff}}$, (b)~cooling rate $\alpha$, (c)~iterations per temperature $N_{\text{iter}}$.
Default values for the other two parameters are held constant in each sweep.}\label{fig:sensitivity}
\end{figure}

We assessed the sensitivity of ParetoEnsembles.jl to three key hyperparameters (rank cutoff $R_{\text{cutoff}}$, cooling rate $\alpha$, and iterations per temperature $N_{\text{iter}}$) by sweeping each while holding the others at default values ($R_{\text{cutoff}} = 8$, $\alpha = 0.90$, $N_{\text{iter}} = 40$) on the Binh--Korn benchmark with five chains and five replicates per configuration (\cref{fig:sensitivity}).
The hypervolume indicator was robust across all tested ranges: rank cutoff values from 2 to 12 produced nearly identical results, with a decline at $R_{\text{cutoff}} = 20$ where the archive retains many low-quality solutions; faster cooling ($\alpha = 0.80$) performed comparably to slower cooling ($\alpha = 0.95$), with the latter showing more variability; and even 10 iterations per temperature level produced competitive fronts, though 20 or more reduced inter-replicate variance.
These results suggest that the default hyperparameters provide a reasonable starting point for most problems, and that users need not invest significant effort in tuning, a practical advantage for the systems biology applications where ParetoEnsembles.jl is primarily intended to be used.

\clearpage
\section{Robustness to model misspecification}\label{sec:si_misspec}
\begin{figure}[ht]
\centering
\includegraphics[width=\textwidth]{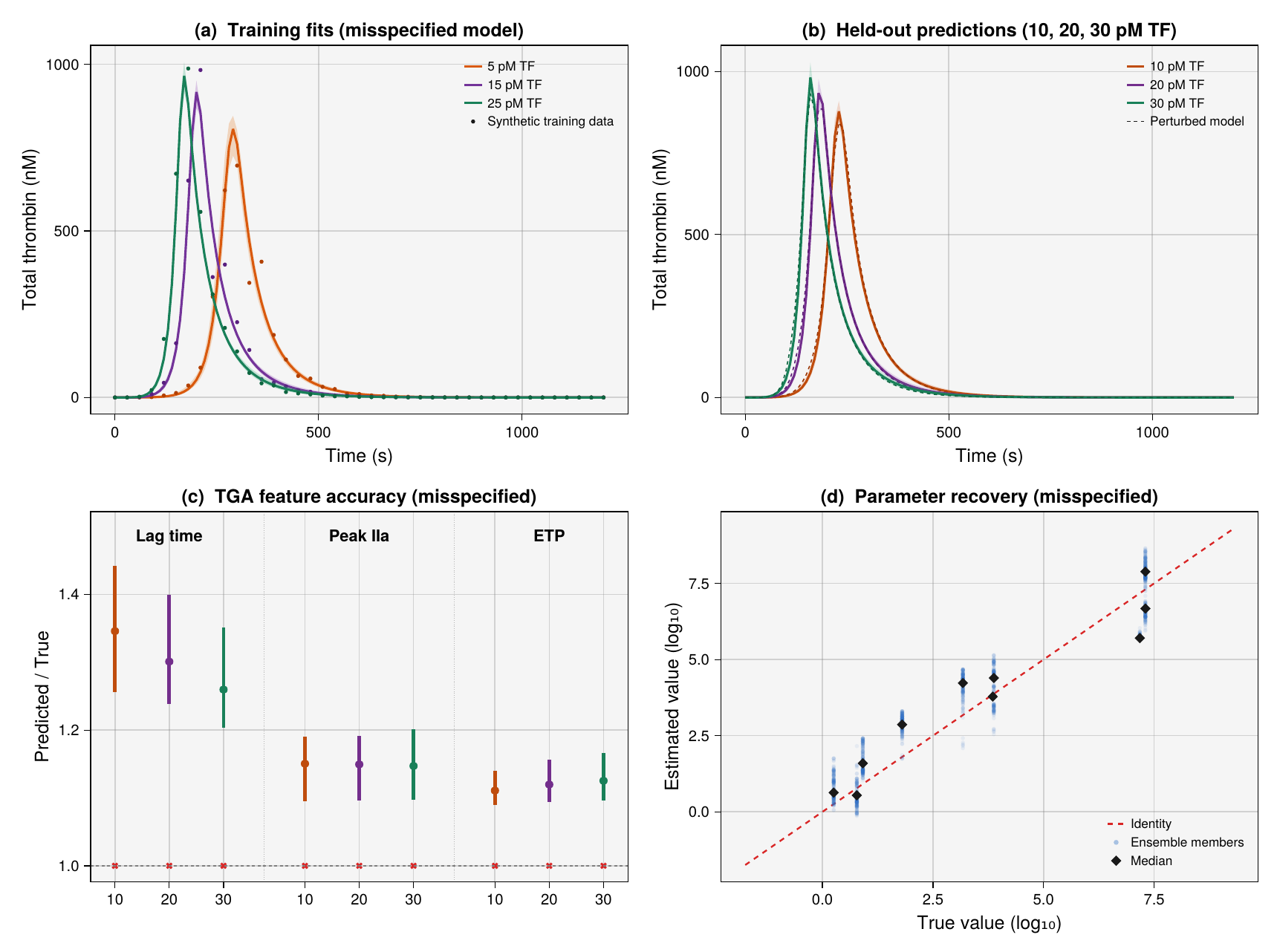}
\caption{Model misspecification study: six fixed rate constants perturbed by $\pm 30\%$ before generating training data, then fitted with the nominal model.
(a)~Training fits at 5, 15, 25~pM TF; the misspecified model still tracks the data.
(b)~Held-out predictions at 10, 20, 30~pM TF; ensemble means (dashed) are within 4--6\% of true trajectories (solid) but 95\% CIs (shaded) do not fully bracket the truth.
(c)~TGA feature accuracy at held-out conditions, normalized to true values; the bias pattern shifts relative to the correct-model case.
(d)~Parameter recovery shows wider scatter and larger biases than the correct model, as expected when the fitting model is structurally approximate.}\label{fig:misspec}
\end{figure}

The coagulation results in the main text use the same model to generate and fit the data, so the fitting model is exactly correct by construction, a setting sometimes called an ``inverse crime'' in the inverse problems literature (\cref{fig:misspec}).
To assess how the ensemble approach performs when this assumption is violated, we repeated the coagulation study with a deliberately misspecified model: six rate constants that are \emph{not} among the ten estimated parameters were perturbed by $\pm 30\%$ from their nominal values before generating the synthetic training data, so that no parameter vector in the fitting model's search space can perfectly reproduce the data.
The perturbed parameters span the initiation (TF=VIIa$+$VII$\to$VIIa), binding (extrinsic Xase and IX activation $K_m$), assembly (prothrombinase on-rate), and inhibition (Xa$+$TFPI and mIIa$+$ATIII) pathways, ensuring that the misspecification affects multiple phases of the coagulation cascade.

Despite the model error, the ensemble still produced thrombin trajectories that closely tracked the training data at all three TF concentrations (\cref{fig:misspec}a), demonstrating that the ten estimated parameters absorbed much of the misspecification through compensatory adjustments.
At the held-out validation conditions (10, 20, and 30~pM TF), the ensemble predicted peak thrombin to within 4--6\% of the true values (\cref{fig:misspec}b), comparable to the 6--7\% error observed with the correct model in the main text.
However, trajectory-level coverage degraded substantially: the 95\% prediction intervals covered the true trajectory at only 31--48\% of time points, compared with higher coverage under the correct model, reflecting the ensemble's inability to account for structural model error through parametric uncertainty alone.

The pattern of TGA feature accuracy also shifted under misspecification (\cref{fig:misspec}c).
Lag time, which was well-covered in the correct-model study, showed a systematic overprediction of 5--9\% with the true value falling outside the 95\% interval at some conditions, while ETP, which was biased under the correct model, was now predicted to within 0.3--2.5\% with good coverage.
Peak thrombin remained biased by 4--6\%, similar to the correct-model case.
These shifts reflect the fact that the estimated parameters compensate for the misspecified fixed parameters differently depending on which aspect of the thrombin curve is being evaluated. They also highlight a fundamental limitation of parametric uncertainty characterization: the ensemble can only explore parameter vectors within a given model structure, and when that structure is approximate, prediction intervals may be miscalibrated even though point predictions remain accurate.
Parameter recovery (\cref{fig:misspec}d) showed wider scatter and larger biases than the correct-model case, as expected since the ``true'' parameter values are no longer optimal under the misspecified model.

These results confirm that the ensemble approach is robust to moderate model misspecification, with predictions degrading gracefully, while revealing the boundary between parametric uncertainty, which the ensemble captures, and structural uncertainty, which requires either model expansion or explicit model-error terms to address.

\clearpage
\section{Code examples}\label{sec:si_code}

\begin{lstlisting}[caption={Binh--Korn benchmark with ParetoEnsembles.jl.},label={lst:binhkorn}]
using ParetoEnsembles

function objective_function(x)
    f = zeros(2, 1)
    f[1] = 4.0 * x[1]^2 + 4.0 * x[2]^2
    f[2] = (x[1] - 5)^2 + (x[2] - 5)^2
    # penalty for constraint violations
    lambda = 100.0
    v1 = 25 - (x[1] - 5)^2 - x[2]^2
    v2 = (x[1] - 8)^2 + (x[2] - 3)^2 - 7.7
    f[1] += lambda * (min(0, v1))^2
    f[2] += lambda * (min(0, v2))^2
    return f
end

neighbor_function(x) = clamp.(
    x .* (1 .+ 0.05 * randn(length(x))),
    [0.0, 0.0], [5.0, 3.0])

acceptance_probability_function(R, T) =
    exp(-R[end] / T)

cooling_function(T) = 0.9 * T

(EC, PC, RA) = estimate_ensemble(
    objective_function, neighbor_function,
    acceptance_probability_function, cooling_function,
    [2.5, 1.5]; rank_cutoff=4.0,
    maximum_number_of_iterations=40, show_trace=false)

pareto_idx = findall(RA .== 0)
\end{lstlisting}

\begin{lstlisting}[caption={Multi-chain parallel execution on the Binh--Korn problem (requires \texttt{julia -t4}).},label={lst:parallel}]
using ParetoEnsembles

# ... (same callbacks as Listing 1) ...

initial_states = [
    [2.5, 1.5], [0.5, 2.5],
    [4.0, 0.5], [1.0, 1.0]
]

(EC, PC, RA) = estimate_ensemble_parallel(
    objective_function, neighbor_function,
    acceptance_probability_function, cooling_function,
    initial_states;
    rank_cutoff=4.0, maximum_number_of_iterations=40,
    show_trace=false, rng_seed=42)

println("Solutions: ", size(EC, 2))
println("Pareto-optimal: ", count(RA .== 0))
\end{lstlisting}

\end{document}